\documentclass[final]{ws-p8-50x6-00}
\usepackage{epsfig}

\begin{document}

\title{Dynamical resummation and damping \\in the O(N) model
\footnote{
T\lowercase{alk presented at} SEWM2000, 
M\lowercase{arseille}, 
F\lowercase{rance}, 14--17 
J\lowercase{une}, 2000.}
}

\author{A.~Jakov\'ac}
\address{Technical University, H-1521 Budafoki \'ut 8, Budapest
\footnote{Present address: 
Theory division, CERN, CH-1211, Geneva 23, Switzerland} 
}

\maketitle

\def\lsi{\raise0.3ex\hbox{$<$\kern-0.75em\raise-1.1ex\hbox{$\sim$}}}
\def\gsi{\raise0.3ex\hbox{$>$\kern-0.75em\raise-1.1ex\hbox{$\sim$}}}
\newcommand{\lsim}{\mathop{\lsi}}
\newcommand{\gsim}{\mathop{\gsi}}

\newcommand{\eqref}[1]{(\ref{#1})}
\newcommand{\cL}{{\cal L}}
\newcommand{\cH}{{\cal H}}
\newcommand{\cA}{{\cal A}}
\newcommand{\sld}{\mbox{$\,\slash$ \hspace{-1.0em}$\nabla$}}
\renewcommand{\d}{\partial}
\newcommand{\dsl}{{\hat\delta}}
\newcommand{\nn}{\nonumber\\}
\newcommand{\tH}{\tilde{H}}
\newcommand{\x}{{\bf x}}
\newcommand{\y}{{\bf y}}
\newcommand{\ph}{\varphi}
\newcommand{\rh}{\varrho}
\newcommand{\bP}{\bar\Phi}
\newcommand{\exv}[1]{\left\langle{#1}\right\rangle}
\newcommand{\exvs}[1]{\langle{#1}\rangle}
\newcommand{\ep}{\varepsilon}
\newcommand{\GeV}{\textrm{~GeV~}}
\newcommand{\q}{{\bf q}}
\renewcommand{\k}{{\bf k}}
\newcommand{\p}{{\bf p}}
\newcommand{\T}{\textrm{T}}
\newcommand{\Tr}{\mathop{\textrm{Tr}}}
\renewcommand{\Im}{\,\textrm{Im}\,}
\renewcommand{\Re}{\,\textrm{Re}\,}

\abstracts{ In this talk I summarize the one loop and higher loop
  calculations of the effective equations of motion of the $O(N)$
  symmetric scalar model in the linear response approximation. At one
  loop one finds essential difference in long time behavior for the
  fields below and above a dynamically generated length scale. A
  partial resummation assuming quasi-particle propagation seems to
  cancel the relevance of this scale.  }

\section{Introduction}

The out of equilibrium behavior of the field theories can play
important role in understanding many physical phenomena, as for
example the cosmological inflation, reheating or some aspects of heavy
ion physics. A possible treatment of these processes is to compute
effective equations of motion (EOM) for the field expectation values
and then solve these equations, most simply by applying one loop
perturbation theory and linear response approximation. These
approximations, however, may not give correct answers in certain
dynamical regions, as calculations in gauge theories show, where
linear response spoils gauge invariance\cite{HTL}, higher loop effects
change the theory completely at the ultra-soft scale\cite{Bodeker}. In
this talk I would like to examine the effects of higher loops on the
dynamical behavior of the $O(N)$ model in linear response
approximation. For details and references c.f. Ref\cite{JPPSz,J}.

What new effects may we expect? In calculation of the imaginary part
the cutting of a higher loop diagram provides more phase space, less
constraint to the incoming momentum. This effect can be important,
when the one loop contribution is small, as in the case of Goldstone
damping in the O(N) model. Here the mass shell constraints send the
internal Goldstone momentum into infinity at one loop resulting in
exponentially suppressed Goldstone damping.

The other expected effect is that while at one loop the internal
particles are stable, higher loops may provide imaginary part for
their propagator. The stability of internal particles leads to long
time memory of the system\cite{BAVH}, their decay, on the other hand,
leads to loss of memory.

In the followings I first summarize the results of one loop linear
response theory, then the results with resummed self energy
propagators.

\section{One loop linear response theory}

The action of the theory
\begin{equation}
  S= \int\left[\frac12(\d\hat\Phi_a)^2 - \frac{m^2}2\hat
  \Phi_a^2 - \frac{\lambda}{24} (\hat\Phi_a^2)^2\right].
\end{equation}
We want to calculate the EOM for the expectation value of the field
$\Phi=\Tr\hat\Phi \equiv \exvs{\hat \Phi}$, where $\rho$ is some
initial density matrix. We apply the operator EOM $\frac{\delta
  S}{\delta\hat\Phi(x)} = 0$ to the decomposition $\hat\Phi=
\Phi+\ph$ (here, by construction, $\exvs{\ph}=0$). Then we take
expectation value and obtain
\begin{equation}
  0 = (\d^2 + m^2 +\frac\lambda6\Phi^2(x) )\Phi(x) + J^{ind}(x),
\label{PEOM}
\end{equation}
where the quantum induced current is
\begin{equation}
  J^{ind}_a(x) = \frac\lambda6 \biggl[\Phi_a(x)\exv{\ph_b^2(x)} +
  2\Phi_b(x)\exv{\ph_b(x)\ph_a(x)} + \exv{\ph_a(x)\ph_b^2(x)}\biggr].
\end{equation}
The expectation values are calculated using real time one loop
perturbation theory (there $\exv{\ph_a\ph_b^2}=0$) in linear response
approximation. We assume moreover that the fluctuations are in
equilibrium. We concentrate on the broken phase where, with proper
choice of the coordinate system, we write
$\Phi_a\to\bP\delta_{a1}+\Phi_a$ with constant $\bP$. The EOM
\eqref{PEOM} determines the value of $\bP$.

In linear response approximation $J^{ind}_a(k) = \Pi^R_{ab}(k)\Phi_b$,
where $\Pi^R$ is the retarded self energy. In the present case it
turns out that the self energy is diagonal $\Pi^R_{ab}=\Pi^R_a
\delta_{ab}$, and
\begin{eqnarray}
  && \Pi_1^R(k) = \frac\lambda2\left[S_1 + \bP^2
    S_{11}(k)\right] + \frac{(N-1)\lambda}6\left[S_i +
    \frac{\bP^2}3 S_{ii}(k)\right]\nn
  && \Pi_i^R(k) = \frac\lambda6\left[S_1+(N+1)S_i\right] +
    \frac\lambda9\bP^2 S_{1i}(k),
\end{eqnarray}
where
\begin{equation}
  S_a=\int\!\frac{d^4q}{(2\pi)^4}\, n(q_0)\rh_a(q),\qquad
  iS_{ab}(x)=\Theta(x_0)\rho_{ab}(x),
\end{equation}
and
\begin{equation}
  \rho_{ab}(k) = \int\!\frac{d^4q}{(2\pi)^4}\,
  \rh_a(q)\rh_b(k-q)(1+n(q_0)+n(k_0-q_0)).
\end{equation}
Here $\rh_a(k)=(2\pi)\ep(k_0)\delta(k^2-m_a^2)$ free spectral function
and $n$ is the Bose-Einstein distribution.

To avoid IR divergences we have to perform (mass) resummation. Here it
is done by using $m_H^2=\frac\lambda3\bP$ and $m_G^2=0$ in the
propagators with the one-loop value of $\bP$.

For a detailed computation see Ref.~\cite{JPPSz}, here I just quote
the main results: {\em 1.)} Different methods (kinetic theory,
perturbation theory, equations for mode functions) give the same
result in one loop linear response approximation. {\em 2.)} Goldstone
theorem is fulfilled in the present approximation scheme taking into
account the quantum corrections to $\bP$ (i.e. $\Pi_i^R(k=0)=0$).
{\em 3.)} The calculations can be done analytically. The damping rates
can be read off from the imaginary part at the mass shell. At high
temperatures
\begin{equation}
  \gamma_1 = (N-1)\frac{\lambda T}{48\pi},\qquad
  \gamma_i=\frac{\lambda m_H^2}{96\pi|\k|} \,n(\frac{m_H^2}{4|\k|}).
\end{equation}
The leading term in damping rate of the Higgs mode is classical (can
be obtained using classical statistical field theory), but for the
Goldstone mode it is classical only for large momenta $|\k|>M\equiv
\frac{m_H^2}{4T}$. For small momenta $|\k|<M$ the Goldstone damping is
exponentially suppressed $\gamma_i\sim e^{-M/|\k|}$ (c.f. also
Ref\cite{PT}.).

\section{Beyond one loop}

Already in the plain one loop case it was necessary to apply some
resummation in order to avoid IR divergences. Similar ideas can be
used to resum self energy diagrams. We add a term to and subtract the
same term from the original Lagrangian
\begin{equation}
  \cL= \cL - \frac12 \int\!d^4y\, \hat \Phi_a(x) P_{ab}(x,y) \hat
  \Phi_b(y) + \frac12 \int\!d^4y\, \hat \Phi_a(x) P_{ab}(x,y) \hat
  \Phi_b(y),
\end{equation}
where $P$ depends also on $\bP$ in the broken phase. We treat the
first term as part of the propagator, the second one as counterterm.
In this way we did not change the physics, at infinite loop order $P$
is irrelevant. At finite loop order, however, the results are
sensitive to the choice of $P$, which sensitivity can be used to
optimize the perturbation theory. We can demand, for example, that the
one loop correction to the self energy (propagator) be zero. There are
two contributions, one comes from a direct calculation with the new
propagator, the other is the counterterm. Their cancellation leads to
a gap equation
\begin{equation}
  \Pi^R(P,\bP) = P,
\label{gapeq}
\end{equation}
where we have denoted the explicit dependence of the self energy on
$P$ (through the propagator) and on the background. In the later
calculations we shall use the resulting $P=\bar P(\bP)$ function.
Since the Lagrangian was symmetric under $O(N)$ rotation where $P$ was
transformed as a tensor, the $\bar P$ solution transforms also as a
tensor under the rotation of the background.  Using this function
instead of $P$ we maintain the $O(N)$ symmetry of the Lagrangian. We
assume in the sequel that we have chosen the coordinate system
properly and $\bar P(\bP)$ is diagonal.

When $P$ is fixed, the calculation goes like in the symmetric phase,
but the propagator changes. We can read off the propagators at finite
temperature from the spectral function as
\begin{equation}
  \begin{array}[c]{ll}
    iG_a^<(k)=n(k_0)\rh_a(k),\quad &
    iG_a^>(k)=(1+n(k_0))\rh_a(k),\cr
    iG_a^c(t,\k)=\Theta(t)\rh_a(t,\k) + iG_a^<(k),\quad &
    iG_a^a(t,\k)=i G_a^>(k)- \Theta(t)\rh_a(t,\k),\cr
  \end{array}
\end{equation}
and the spectral function can be expressed in the present case as
\begin{equation}
  \rh_a(k) = \frac{-2\Im\bar P_a}{(p^2-m_a^2-\Re\bar P_a)^2 + \Im \bar
  P^2}.
\end{equation}
These relations make the gap equation \eqref{gapeq} explicit. 

To have an analytical solution we have to make some assumptions. We
use Breit-Wigner approximation (assuming pole dominance), i.e. we
approximate the true spectral function as
\begin{equation}
  \rh(k)\approx \frac\pi{\omega_\k}\biggl(
  \delta_{\gamma_\k}(k_0-\omega_\k) -
  \delta_{\gamma_\k}(k_0+\omega_\k) \biggr),
\end{equation}
where $\delta_\gamma(\omega) =
\frac1\pi\frac\gamma{\omega^2+\gamma^2}$ smeared delta-function.
Passing by the calculations (c.f. \cite{J}) I summarize the changes
compared to the $\gamma=0$ case: {\em 1.)} Instead of Landau
prescription $k_0\to k_0+i\ep$ we find $k_0\to k_0+i\hat\gamma$ in
$S_{ab}$, where $\hat\gamma = \gamma_{a\k}+\gamma_{b\k}$. {\em 2.)}
Instead of strict energy conservation the energy is conserved by
$\delta_{\hat\gamma}$ in calculation of the imaginary part. {\em 3.)}
For low momenta ($|\k|< \gamma_1$ for Higgs and $|\k|<m_H$ for
Goldstones) both the Higgs and Goldstone fields have imaginary part
proportional to $k_0$
\begin{equation}
  \Im\Pi^R_a(k) = -\eta_a k_0,
\end{equation}
where $\eta_1\sim T\lambda^2\log\lambda$ and $\eta_i\sim m_H$.

This latter point yields finite on-shell damping rate for the
Goldstone modes, showing that the one loop result was not reliable as
expected in the Introduction. It also means that in the effective
equation of motion a term $\sim\dot\Phi$ appears instead of the
integral over the past. That is the loss-of-memory effect indicated in
the Introduction.

\section{Conclusions}

We have computed the effective EOM for the $O(N)$ model in the linear
response approximation at one loop level and with self energy
resummation. At one loop we find that the Higgs dynamics in the
leading temperature order is consistent with the classical
expectations, while for the Goldstone we obtain exponentially small
damping rate for momenta $|\k|<M=m_H^2/4T$
\begin{equation}
  \gamma_i^{1-loop}\sim e^{-M/|\k|}.
\end{equation}
To go beyond one loop we have performed self energy resummation
formulated in gap equations. For the solution we have used
Breit-Wigner approximation, which have modified in the result the Landau
prescription (now $k_0\to k_0 + i\hat\gamma$) and have resulted in a
broadened mass shell for the intermediate particles. As a consequence
we have found that for low momenta
\begin{equation}
  \Pi^R_a \sim -\eta_a \d_t
\end{equation}
for both the Goldstone and Higgs fields. Therefore the Goldstone
on-shell damping rate is finite, and we can describe the dynamics of
low momentum fields with a differential equation without long time
memory kernels. 

There can be also other effects which can modify this statement, first
of all the ones coming from the running of the coupling constant. On
the other hand similar considerations may be applicable for other
theories (e.g. gauge theories) as well.

\end{document}